\documentclass[10pt]{sigplanconf}

\usepackage{xspace}
\usepackage{enumitem}
\usepackage{tabularx}
\usepackage{listings}
\usepackage{color}
\usepackage{url}
\usepackage{underscore}
\usepackage{array}
\usepackage{tabu}
\usepackage{graphicx}

\definecolor{grey}{rgb}{0.5,0.5,0.5}

\newcommand{\name}{Falcon\xspace}
\newcommand{\ntilde}{\raise.17ex\hbox{$\scriptstyle\sim$}}

\newcommand{\rcode}{register code\xspace}
\newcommand{\Rcode}{Register code\xspace}

\newcommand{\pcode}[1]{{\ttfamily\scriptsize{}#1}}

\newcommand{\rstack}[1]{$\langle{}#1\rangle{}$} 
\newcommand{\stackop}[2]{$\langle{}#1\rangle{}\rightarrow\langle{}#2\rangle{}$}

\begin{document}
%\conferenceinfo{DLS '13}{October 28, 2013, Indianapolis, Indiana, USA}
%\copyrightyear{2013}
\title{How fast can we make interpreted Python?}

\authorinfo{Russell Power\and Alex Rubinsteyn}
{New York University}
{\{power,alexr\}@cs.nyu.edu}

\maketitle

\abstract

Python is a popular dynamic language with a large part of its appeal coming
from powerful libraries and extension modules.  These augment the language and
make it a productive environment for a wide variety of tasks, ranging from web
development (Django) to numerical analysis (NumPy).

Unfortunately, Python's performance is quite poor when compared to modern
implementations of languages such as Lua and JavaScript.  Why does Python lag so
far behind these other languages? As we show, the very same API and extension
libraries that make Python a powerful language also make it very difficult to
efficiently execute.

Given that we want to retain access to the great extension libraries that
already exist for Python, how fast can we make it?  To evaluate this, we
designed and implemented \name, a high-performance bytecode interpreter fully
compatible with the standard CPython interpreter. \name applies a number of well
known optimizations and introduces several new techniques to speed up execution
of Python bytecode.  In our evaluation, we found \name an average of 25\%
faster than the standard Python interpreter on most benchmarks and
in some cases about 2.5X faster.

\section{Introduction}

Python is popular programming language, with a long history and an active
development community.  A major driver of Python's popularity is the diverse
ecosystem of libraries and extension modules which make it easy to do almost
anything, from writing web-servers to numerical computing.  But despite
significant effort by Python's developers, the performance of Python's
interpreter still lags far behind implementations of languages such as Lua and
JavaScript.  

What differentiates Python from ``faster'' languages? Obviously, implementation
choices (JIT vs. interpreter) can have a dramatic effect on performance. In the
case of Python, however, the landscape of choices is severely constrained by
the very same API that makes it easy to extend. The standard interpreter for
Python (called CPython) exposes a low-level API (the Python C
API ~\cite{python-api})
 which allows for building extension libraries and for
embedding the interpreter in other programs.  The Python C API allows access
to almost every aspect of the interpreter, including inspecting the current
interpreter state (what threads are running, function stacks, etc..) and pulling
apart the representation of various object types (integer, float, list,
dictionary, or user-defined).  For performance reasons, many Python libraries
are written in C or another compiled language, and interface to Python via the
C API.

In an ideal world, any implementation of Python's semi-formal
specification~\cite{python} would be interchangeable with the CPython
implementation.  Unfortunately, to avoid breaking libraries, an alternative
implementation must also support the full C API. While the size of the C API
(\ntilde 700 functions) is burdensome, what really makes it problematic is the
degree to which it exposes the internal memory layout and behavior of Python
objects.  As a result of this many Python extensions have become intimately
coupled to the current implementation of the CPython interpreter. For instance,
modifying the layout of the basic object format (for example, to use less
memory) breaks even source level compatibility with existing extensions.

The value of these extensions to Python is hard to overstate. Python already has
a fast JIT compiler in the form of PyPy~\cite{bolz09}, but is has
not seen widespread adoption.  This is, to a large extent, due to the lack of
support for existing CPython extension libraries.  Replacing these libraries is
not a simple undertaking; NumPy~\cite{numpy} alone consists of almost 100k lines
of C source, and the SciPy libraries which build upon it are another 500k lines.

%If we accept the restriction of binary compatibility with existing extension
%libraries, there are still a few techniques that can be used to create a faster interpreter.
To evaluate how much we can improve over CPython (without breaking existing extension modules), 
we developed an alternative Python virtual machine called \name. 
\name converts CPython's stack-oriented bytecode into a register-based format and then
performs optimizations to remove unnecessary operations and
occasionally elide type checks.  \name's register bytecode is then executed by a 
threaded interpreter, which attempts to accelerate common code patterns using attribute lookup caching
and register tagging.

Overall, we found that: 
\begin{itemize}
	\item A combination of register conversion, simple bytecode optimization, and a virtual machine with threaded dispatch result in an average 25\% speedup over CPython. On certain benchmarks, \name was up to 2.5X faster. 
	\item Just-in-time register conversion and optimization are fast enough to run online for every function. This means that a system like \name can accelerate any code without the need for profiling and compilation heuristics. 
\end{itemize}

\section{Overview}

%\name is implemented as a Python extension module which runs within CPython.
%A user can indicate which function they want to run inside \name by wrapping it
%with the decorator function \pcode{@falcon.wrap}. When a decorated function is
%first called, \name takes over evaluation.

\name does not replace the standard CPython interpreter, but rather runs inside
of it.  A user-defined function can be marked for execution by \name with
the decorator \pcode{@falcon}. When a decorated function is called, \name
translates that function's stack bytecode into \name's more compact \rcode. This \rcode
is then optimized to remove redundant computations and decrease the
number of registers needed by a function. The optimized \rcode is then passed
on to \name's virtual machine for evaluation.

% When a function executing within \name calls some other function, \name
% attempts to also translate and  execute the function being called. If this is
% not possible, \name temporarily returns control back to CPython.
For example, consider executing the following function, which adds its two
inputs, assigns their sum to a local variable, and then returns that variable:

\begin{figure}[h!]
\begin{lstlisting}
def add(x,y):
  z = x + y
  return z
\end{lstlisting}
\caption{\label{fig:add-python} Python function that adds two inputs}
\end{figure}

When Python first encounters this code, it is compiled into the following
\emph{bytecode}.

\begin{figure}[h!]
\begin{lstlisting}
  LOAD_FAST        (x)
  LOAD_FAST        (y)
  BINARY_ADD          
  STORE_FAST       (z)
  LOAD_FAST        (z)
  RETURN_VALUE
\end{lstlisting}

\caption{\label{fig:add-stack} Python stack code that adds two inputs}
\end{figure}

Each operation in Python's bytecode implicitly interacts with a value stack.
Python first pushes the values of the local variables \pcode{x} and \pcode{y}
onto the stack. The instruction \pcode{BINARY_ADD} then takes these two values,
adds them, and pushes this new result value onto the stack. Where did the values
of the local variables come from? In addition to the stack, Python's virtual
machine maintains a distinct array of values for named local variables. The
numerical arguments attached to the \pcode{LOAD_FAST} and \pcode{STORE_FAST}
instructions indicate which local variable is being loaded or stored.

Even from this simple example, we can see that Python bytecode burdens the
virtual machine with a great deal of wasteful stack manipulations. Could we get
better performance if we did away with the stack and instead only used the array
of local values? This is the essence of a \emph{register bytecode}. Every instruction
explicitly labels which registers (local variable slots) it reads from and to
which register it writes its result.

Translated into \rcode, the above example would look like:

\begin{figure}[h!]
\begin{lstlisting}
  r2 = BINARY_ADD(r0, r1)
  RETURN_VALUE r2
\end{lstlisting}
\caption{\label{fig:add-register} \Rcode for adding two inputs}
\end{figure}

When converted into \rcode, the local variables \pcode{x}, \pcode{y} and
\pcode{z} are represented by the registers \pcode{r0}, \pcode{r1} and
\pcode{r2}. Since the source and destination registers are part of each
instruction, it is possible to express this function using only two
instructions.

The downside to \name's \rcode format (like any \rcode) is that each instruction
must be larger to make room for register arguments. There are two advantages to
\rcode which make the space increase worthwhile.

The first is a potential for reducing the time spent in virtual machine dispatch
by reducing the number of instructions that must be executed. Previous research
has verified that switching from a stack-based virtual machine to a register
machines can improve performance~\cite{davis03, showdown08}.

Additionally, it is much easier to write optimizations for a \rcode. The reason
for this is that when every instruction implicitly affects a stack, program
analyses and optimizations must track these side effects in some form of virtual
stack. A \rcode, on the other hand, admits the expression of more compact
optimization implementations (section~\ref{section:opt}), since instructions
only depend on each other through explicit flows of data along named registers.

\section{Compiler}

The \name compiler is structured as a series of passes, each of which
modifies the \rcode in some way.  To illustrate the behavior of each pass 
we will use a simple example function called \pcode{count_threshold} - which
counts the number of elements in a list below a given threshold:

\begin{lstlisting}
def count_threshold(x,t): 
  return sum([xi < t for xi in x])
\end{lstlisting}

For \pcode{count_threshold} Python generates the following stack
bytecode (Figure~\ref{fig:stackcode}):

\begin{figure}[h!]
\begin{lstlisting}
    LOAD_GLOBAL       (sum)
    BUILD_LIST        
    LOAD_FAST         (x)
    GET_ITER         
10: FOR_ITER          (to 31)
    STORE_FAST        (xi)
    LOAD_FAST         (xi)
    LOAD_FAST         (t)
    COMPARE_OP        (<)
    LIST_APPEND       
    JUMP_ABSOLUTE     10
31: CALL_FUNCTION     
    RETURN_VALUE 
\end{lstlisting}
\caption{Python stack machine bytecode \label{fig:stackcode}}
\end{figure}

Before we can do anything else, we need to convert our original stack machine
bytecode to the equivalent \rcode.

\subsection{Stack-to-register conversion}

To convert from Python stack code to register code \name uses abstract
interpretation~\cite{abstract-interpretation}. In \name this takes the form of a
``virtual stack" which stores register names instead of values.    \name steps
through a function's stack operations, evaluates the effect each has on the
virtual stack, and emit an equivalent register machine operation.

\subsubsection*{Handling control flow} 

For straight-line code this process is fairly easy; most Python instructions
have a fairly straightforward effect on the stack.  But what happens when we
encounter a branch?  We need to properly simulate both execution paths.
To handle this situation, we must make a copy of our virtual stack, and evaluate
both sides of the branch.

With branches come merge points; places where two or more branches of execution
come together. Each thread of control flow might have assigned different
register names to each stack position. To handle this situation \name inserts
rename instructions before merge points, ensuring that all incoming register
stacks are compatible with each other.  (This is the same mechanism employed by
compilers which use static single assignment form (SSA)\cite{ssa} to resolve
$\phi$-nodes.)

% Curiously, this exact issue of
% conflicting names as well as  the method for resolving the conflict, is
% identical to that found in compilers which use static single assignment
% (SSA)\cite{ssa} for optimization.
%% ---> 
%% IT'S NOT REALLY SSA SINCE WE DON'T GIVE A UNIQUE NAME TO EVERY ASSIGNMENT
%% TO A LOCAL...IT SORT OF RESEMBLES THE CODE YOU GET WHEN YOU NAIVELY 
%% TRANSLATE OUT OF SSA. WE *COULD* HANDLE THE MERGE POINTS BY EXPLICITLY 
%% LABELING FLOWS BUT IT DOESN'T SEEM WORTH IT.
%
% We're not claiming that it's SSA; merely that the method of handling merge
% points turns out to be the same.

\subsubsection*{Example conversion} 
Let's walk through how this works for the  example stack code above
(figure~\ref{fig:stackcode}).

First we find the value of the function ``sum'' using the \pcode{LOAD_GLOBAL}
instruction. In the CPython interpreter, \pcode{LOAD_GLOBAL} looks up a
particular name in the dictionary of global values and pushes that value onto
the stack. Since the set of literal names used in a function is known at compile
time, the instruction can simply reference the index of the string ``sum'' in a
table of constant names. The equivalent register machine instruction assigns the
global value to a fresh register (in this case \pcode{r4}).  For brevity, the
``stack'' column in the listings below will show just the register number for
each instruction.

\begin{figure}[h!]
	\begin{tabularx}{8.45cm} { |l|l|X|}
  \hline
  \textbf{Python}  & \textbf{Falcon}  & \textbf{Stack} \\ \hline
  \pcode{LOAD_GLOBAL 0} & \pcode{r4 = LOAD_GLOBAL 0} & \stackop{}{4}\\ \hline
  \end{tabularx}
\end{figure}

The effect of this operation on the virtual stack is to push the register
\pcode{r4} on top. When a later operation consumes inputs off the stack, it will
be correctly wired to use \pcode{r4} as an argument.  

\pcode{BUILD_LIST} constructs an empty list to contain the results.  We create a
new register \pcode{r5} and push it onto the stack.

\begin{figure}[h!]
  \begin{tabularx}{8.45cm} { |l|l|X|}
  \hline
  \textbf{Python}  & \textbf{Falcon}  & \textbf{Stack} \\ \hline
  \pcode{BUILD_LIST 0} & \pcode{r5 = BUILD_LIST 0} & \stackop{4}{5,4}\\ \hline
  \end{tabularx}
\end{figure}

Python has special operations to load and store local variables and to load
constants.  Rather than implement these instructions directly, we can
alias these variables to specially designated register names, which simplifies our
code and reduces the number of instructions needed.

\begin{figure}[h!]
	\begin{tabularx}{8.45cm} { |l|l|X|}
  \hline
  \textbf{Python}  & \textbf{Falcon}  & \textbf{Stack} \\ \hline
  \pcode{LOAD_FAST 0 (x)} &  & \stackop{5,4}{1,5,4} \\ \hline
  \end{tabularx}
\end{figure}

Register \emph{r1} is aliased to the local variable \emph{x}.  Therefore for the
\pcode{LOAD_FAST} operation here, we don't need to generate a \name instruction,
and can instead simply push \emph{r1} onto our virtual stack.

\pcode{GET_ITER} pops a sequence off of the stack and pushes back an iterator
for the sequence.

\begin{figure}[h!]
  \begin{tabularx}{8.45cm} { |l|l|X|}
  \hline
  \textbf{Python}  & \textbf{Falcon}  & \textbf{Stack} \\ \hline
  \pcode{GET_ITER} & \pcode{r6 = GET_ITER(r1)} & \stackop{1,5,4}{6,5,4}\\
  \hline
  \end{tabularx}
\end{figure}

\pcode{FOR_ITER} is a branch instruction.  It either pushes the next element in
the iterator onto the stack and falls-through to the next instruction, or pops
the iterator off the stack and jumps to the other side of the loop.

\begin{figure}[h!]
  \begin{tabularx}{8.45cm} { |l|l|X|}
  \hline
  \textbf{Python}  & \textbf{Falcon}  & \textbf{Stack} \\ \hline
  \pcode{FOR_ITER} & \pcode{r7 = FOR_ITER(r6)} & \stackop{6,5,4}{7,6,5,4} \newline or \rstack{5,4} \\
  \hline
  \end{tabularx}
\end{figure}

One branch of the \pcode{FOR_ITER} instruction takes us into inner loop, which
continues until the iterator is exhausted:

\begin{figure}[h!]
  \begin{tabularx}{8.45cm} { |l|l|l|}
  \hline
  \textbf{Python}  & \textbf{Falcon}  & \textbf{Stack} \\ \hline
  \pcode{STORE_FAST (xi)} & \pcode{r3 = r7} & \stackop{7,6,5,4}{6,5,4} \\
  \pcode{LOAD_FAST (xi)} & & \stackop{6,5,4}{3,6,5,4} \\
  \pcode{LOAD_FAST (t)} & & \stackop{3,6,5,4}{2,3,6,5,4} \\
  \pcode{COMPARE_OP} & \pcode{r8 = r3 > r2} & \stackop{2,3,6,5,4}{8,6,5,4} \\
  \pcode{LIST_APPEND} & \pcode{APPEND(r5, r8)} & \stackop{8,6,5,4}{6,5,4} \\
  \pcode{JUMP_ABSOLUTE} & \pcode{JUMP_ABSOLUTE} & \rstack{6,5,4} \\ \hline
  \end{tabularx}
\end{figure}

The behavior of the \pcode{LIST_APPEND} instruction here might look somewhat
surprising; it appears to ``peek into'' the stack to find \emph{r5}.  This
special behavior is unique to the \pcode{LIST_APPEND} instruction, and likely is
a result of past performance tuning in the CPython interpreter (building lists
is a very common operation in Python).

And the other branch takes us to our function's epilogue:

\begin{figure}[h!]
  \begin{tabularx}{8.45cm} { |l|l|X|}
  \hline
  \textbf{Python}  & \textbf{Falcon}  & \textbf{Stack} \\ \hline
  \pcode{CALL_FUNCTION (sum)} & \pcode{r9 = sum(r4)} & \stackop{5,4}{6} \\
  \pcode{RETURN_VALUE} & \pcode{RETURN_VALUE(r9)} & \stackop{6}{} \\ \hline
  \end{tabularx}
\end{figure}

\subsubsection*{Operations with dynamic stack effects} In the above example, the
effect of each instruction on the stack was known statically.  It turns out that
this is the case for \emph{almost} all Python instructions.  In fact, only one
operation (\pcode{END_FINALLY}) has a stack effect that must be determined at
runtime.  This instruction appears in functions which have a \pcode{try\ldots{}finally}
block, and determines whether a caught exception should be re-raised.  While it
is possible to handle such an instruction dynamically (by inserting branches in
the generated code for each possible stack effect), we chose a much simpler
option - we simply do not compile functions containing this instruction.
Instead these functions are evaluated using the existing Python interpreter.  As
this instruction is relatively rare, (occurring in only 4\% of the functions in
the Python standard library), and is almost never found in performance sensitive
code, the cost of not supporting it is minimal.

\subsection{Bytecode Optimizations}
\label{section:opt}

After the stack to register pass, the bytecode for \\ \pcode{count_threshold}
now looks like Figure~\ref{fig:unoptimized}.

\begin{figure}[h!]
\begin{lstlisting}
bb_0:
  r4 = LOAD_GLOBAL (sum)
  r5 = BUILD_LIST
  r6 = GET_ITER(r1) -> bb_10
bb_10: 
  r7 = FOR_ITER(r6) -> bb_13,bb_31
bb_13: 
  r3 = r7
  r8 = COMPARE_OP(r3, r2)
  LIST_APPEND(r5, r8)
  JUMP_ABSOLUTE() -> bb_10
bb_31: 
  r9 = CALL_FUNCTION(r5, r4)
  RETURN_VALUE(r9)
\end{lstlisting}
\caption{Unoptimized register code \label{fig:unoptimized}}
\end{figure}

Note that rather than using positions in the code for jump targets, \name splits
up code into basic blocks (indicated with the \pcode{bb_*} prefix). This change
has no effect on the code that ultimately gets run by the virtual machine but
greatly simplifies the implementation of optimization passes.

The register machine bytecode emitted from the stack-to-register pass tends to
be sub-optimal: it uses too many registers and often contains redundant loads
used to emulate the effect of stack operations.  To improve performance, we
perform a number of optimizations that remove these redundant operations and
improve register usage.

The advantage of switching to a \rcode becomes clear at this point; we can apply
known optimization techniques to our unoptimized \rcode with almost no
modification.  The optimizations used in \name are common to most compilers; we
briefly describe them and their application to \name here.

\subsubsection*{Copy Propagation} Whenever a value is copied between registers
(e.g. \pcode{r3 = r7}), it is possible to change later uses of the target
register (\pcode{r3}) to instead use the original source (\pcode{r7}).
In the code from Figure~\ref{fig:unoptimized}, copy propagation changes 
``\pcode{r8 = COMPARE_OP(r3, r2)}'' into ``\pcode{r8 = COMPARE_OP(r7, r2)}''.  By itself, this
optimization will not improve performance.  Instead, it enables other
optimizations to remove useless instructions and to reuse unoccupied registers.

\subsubsection*{Dead Code Elimination} If the value contained in a register is
never used in a program (likely to occur after copy propagation), it may be
possible to delete the instruction which ``created'' that register. Instructions
which lack side effects (simple moves between registers) are safe to
delete, whereas instructions which may run user-defined code (such as
\pcode{BINARY_ADD}) must be preserved even if their result goes unused. Once
copy propagation is applied to the code above, the register \pcode{r3} is never
used. Thus, the move instruction ``\pcode{r3 = r7}'' gets deleted by dead code
elimination.

\subsubsection*{Register Renaming} Even if a register is used at some point in
the program, it might not necessarily be ``alive'' for the entire duration of a
function's execution. When two registers have non-overlapping live ranges, it
may be possible to keep just one of them and replace all uses of the other
register. This reduces the total number of registers needed to run a function,
saving memory and giving a slight performance boost.

\subsubsection*{Register code after optimization}

After \name's optimizations are applied to the bytecode
in~\ref{fig:unoptimized}, extraneous store instructions (such as \pcode{r3 =
r7}) are removed. Furthermore, register renaming causes the registers \pcode{r7}
and \pcode{r4} to be used repeatedly in place of several other registers.

The optimized register bytecode achieves a greater instruction density, compared
with the original stack code.  Optimization in general reduces the number of
instructions by 30\%, with a similar improvement in performance.

\begin{figure}[h]
\begin{lstlisting}
 bb_0: 
  r4 = LOAD_GLOBAL(sum)
  r5 = BUILD_LIST()
  r6 = GET_ITER(r1) -> bb_10
bb_10: 
  r7 = FOR_ITER(r6) -> bb_13,bb_31
bb_13: 
  r7 = COMPARE_OP(r7, r2)
  LIST_APPEND(r5, r7)
  JUMP_ABSOLUTE() -> bb_10
bb_31: 
  r4 = CALL_FUNCTION[1](r5, r4)
  RETURN_VALUE(r4) -> 
\end{lstlisting}
\caption{Optimized register code \label{fig:optimized}}
\end{figure}

\subsubsection*{Difficulty of Optimizing Python Code} It would be desirable to
run even more compiler optimizations such as invariant code motion and common
sub-expression elimination. Unfortunately \emph{these are not valid when
applied to Python bytecode}. The reason these optimizations are invalid is that
almost any Python operation might trigger the execution of user-defined code
with unrestricted side effects. For example, it might be tempting to treat the
second \pcode{BINARY_ADD} in the following example as redundant.

\begin{lstlisting}
  r3 = BINARY_ADD r1, r2 
  r4 = BINARY_ADD r1, r2
\end{lstlisting}

However, due to the possibility of encountering an overloaded \pcode{__add__}
method, no assumptions can be made about the behavior of \pcode{BINARY_ADD}. In
the general absence of type information, almost every instruction must be
treated as \pcode{INVOKE_ARBITRARY_METHOD}.

\section{Virtual Machine} After compilation, the \rcode is passed over to the
virtual machine to evaluate.  \name uses a 3 main techniques (token-threading,
tagged-registers and lookup hints) to try and improve the dispatch performance,
which we cover in this section.  

\subsection*{Token-threading} 
The common, straightforward approach (used by
Python 2.*) to writing a bytecode interpreter is to use a switch and a while
loop:
 
 \begin{figure}[h!]
\begin{lstlisting}
Code* ip = instructions;
while(1) {
  switch(ip->opcode) { 
    case BINARY_ADD:
      Add(ip->reg[0], ip->reg[1], ip->reg[2]);
      break;
    case BINARY_SUBTRACT:
      Sub(ip->reg[0], ip->reg[1], ip->reg[2]);
      break;
  }
  ++ip;
}
\end{lstlisting}
\caption{\label{fig:switch-dispatch} Switch dispatch}
\end{figure}

Most compilers will generate an efficient jump-table based dispatch for this
switch statement. The problem with this style of dispatch is that it will not make effective use of the
branch prediction unit on a CPU.  Since every instruction is dispatched from the
top of the switch statement, the CPU is unable to effectively determine which
instructions tend to follow others. This leads to pipeline stalls and
poor performance. Fortunately, there is an easy way to improve on this.

Token-threading is a technique for improving the performance of switch based
interpreters.  The basic idea is to ``inline" the behavior of our switch
statement at the end of every instruction. This requires a compiler which
supports labels as values~\cite{gcc-labels-as-values} (available in most C
compilers,  with the Microsoft C compiler being a notable exception).

\begin{figure}[h!]
\begin{lstlisting}
jump_table = { &&BINARY_ADD, 
               &&BINARY_SUBTRACT, ... };
BINARY_ADD:
  Add(ip->reg[0], ip->reg[1], ip->reg[2]);
  goto jump_table[(++ip)->opcode];
BINARY_SUBTRACT:
  Sub(ip->reg[0], ip->reg[1], ip->reg[2]);
  goto jump_table[(++ip)->opcode];
\end{lstlisting}
\caption{\label{fig:token-dispatch} Token threading}
\end{figure}

By inlining the jump table lookup we replace the single difficult to predict
branch with many, more predictable branches.  For example, if \pcode{BINARY_ADD}
is always followed by \pcode{BINARY_SUBSCR} in a certain loop, the processor
will be able to accurately predict the branch and avoid stalling.  Token
threading was recently added to the Python 3.1
interpreter\cite{python-token-threading}.

We can go one step further, and modify our bytecode to contain the actual
address of the handler for each instruction.  This results in
\emph{direct-threading}~\cite{direct-threading}.

\begin{figure}[h!]
\begin{lstlisting}
foreach (instr in function) {
  instr.handler = jump_table[ip.opcode]
}

BINARY_ADD:
  Add(ip->reg[0], ip->reg[1], ip->reg[2]);
  goto (++ip)->handler;
\end{lstlisting}
\caption{\label{fig:direct-threading} Direct threading}
\end{figure}

Direct threading increases the size of each instruction while removing a lookup
into the jump table.  We implemented both token and direct threading for \name.
Token threading seems to provide a modest (\ntilde{}5\%) performance improvement over
switch based dispatch.

\subsection*{Tagged Registers} The default Python object format (PyObject) is
inefficient.  A simple word-size integer requires 3 words of space, and must be
dereferenced to get the actual value.  For numerically intensive code, the cost
of this indirection can dominate the interpreter runtime.

This overhead can be reduced by using a more efficient object format.  In
general, changing the object format would break compatibility with existing
code, but here the \name's use of registers proves to be very convenient.  As
long as a value is in a register, we can store it in whatever format is most
efficient.  Only when a value has to be handed to the Python API (or other
external call) do we need to convert it back to the normal Python object format.

For our tests, we chose a simple tagged integer format.  Integer tagging takes
advantage of the fact that object pointers are always \emph{aligned} in memory
to word boundaries; the least significant 2 bits are always zero.  We can
therefore use the least significant bit of a register to indicate whether it is
storing an integer value or a pointer.  If it is storing an integer, we shift
the register right one bit to obtain the value.  Boland~\cite{bolandmemory}
provides detailed descriptions of different tagged representations and their
cost/benefits.

A simplified example of how tagged registers are implemented is show in
Figure~\ref{fig:tagged-register} .

\begin{figure}[h!]
\begin{lstlisting}
struct Register {
  union {
    int int_val;
    PyObject* py_val;
  };
  
  bool is_int() { return int_val & 1; }
  int as_int() { return int_val >> 1; }
  
  // Mask bottom 2 bits out
  PyObject* as_obj() { 
    return py_val & 0xfffffffc; 
  }
};
\end{lstlisting}
\caption{\label{fig:tagged-register}Tagged register format}
\end{figure}

\subsection*{Lookup Hints} 

Attribute lookups (e.g. \pcode{myobj.foo}) are handled by the \pcode{LOAD_ATTR}
instruction and account for a significant portion of the time spent in the Python interpreter. 
Just-in-time compilers for many languages accelerate method lookups using polymorphic inline
caching~\cite{polymorphic-inline-caching} (PIC) and shadow classes.
In many instances, these can replace expensive dictionary (hash-map) with direct
pointer offsets (making them effectively no more costly than a \pcode{struct} in
C).  Unfortunately, this technique is difficult to employ in the context of
\name for a few reasons:
\begin{itemize}
\item \textbf{Fixed object format}. Shadow classes require control over how objects are laid out
in memory, which would break our compatibility goal.

\item \textbf{Complex lookup behavior}. Python provides a great deal of
flexibility to application programmers in choosing how attribute lookup is
performed.  The builtin behavior for resolving lookups is similar to that found
in most languages (first check the object dictionary, then the class dictionary,
then parent classes\ldots).  In addition to this, Python offers an unusual
degree of flexibility for programmers in the form of \emph{accessor methods}.
These allow specifying what happens if an attribute is not found (using the
\pcode{\_\_getattr\_\_} method) or even completely override the normal lookup
process (the \pcode{\_\_getattribute\_\_} method).  What's more, these methods
can be added to a class after objects have been created, which creates
complications for mechanisms like shadow-classes, which expect lookup behavior
to remain consistent.

\item \textbf{Attribute lookup is hidden by the bytecode}. Making matters worse is that the
bytecode generated by Python does not explicitly check whether methods such as
\pcode{\_\_getattr\_\_} have been defined.  Instead, the \pcode{LOAD_ATTR}
instruction is expected to implicitly perform checks for the various accessor
functions for every lookup.
\end{itemize}
One general way of handling such complex behavior is to use a traces within
a just-in-time compiler~\cite{dynamo}.  Unfortunately, this would
increase the complexity of \name by an order of magnitude; while it may be an
appropriate choice in the future, we were interested in determining whether a
simpler approach might be effective.

\name uses a variant on PIC, which we call ``lookup hints''.  As the name
suggests, these provide a guess for where an attribute can be found.
A hint records the location where an attribute was found the last time the
instruction was run.  A hint can indicate that the attribute was in the instance
dictionary an object, or as an attribute of parent class.  When a hint is found,
the location specified by the hint is checked first before the normal lookup
traversal is performed.  If the hint matches, the resulting value is returned
immediately; otherwise the full lookup procedure is performed and a new hint
generated.

The benefit of hints depends greatly on the application being executed.  Code
that references lots of attributes in a consistent manner can see a large
improvement in performance; in general we observed a \ntilde5\% improvement for
most of our benchmarks.

\begin{figure}[h!]
\begin{lstlisting}
void LoadAttr(RegOp op, Evaluator* eval) {
  // Load the hint for this instruction.
  LookupHint h = eval->hint[op.hint_pos];
  PyObject* obj_klass = get_class(op.reg[0]);
  PyObject* obj_dict = get_dict(op.reg[0]);
  PyObject* key = op.reg[1];
  
  // If we previously found our attribute in
  // our instance dictionary, look there again.
  if (h.dict == obj_dict &&
      h.dict_size == obj_dict->size &&
      h.klass == obj_klass &&
      obj_dict.keys[h.offset] == key) {
      return obj_dict.values[h.offset];   
  }
  // no hint, normal path
  ...
}
\end{lstlisting}
\caption{\label{fig:lookup-hints}Simplified implementation lookup hints}
\end{figure}

\section{Implementation}

\name is implemented in C++ as a Python extension module and is compatible with
Python 2.6 through 3.3.  Building on top of the existing Python interpreter
means that \name takes a relatively small amount of code to implement; the
entire package is only 3000 lines of code, evenly split between the
compiler/optimizer and the interpreter.

The method used to implement the \name VM is somewhat unusual.  A typical VM
interpreter is structured as a single method containing the code to handle each
type of instruction.  The approach we took with \name was implemented as a number
of C++ classes, one for each Python opcode.  We use the compiler function
attributes to force the inlining of the code for each opcode into the main
interpreter dispatch loop.  We have found this technique to be very effective,
allowing a clean separation of code without sacrificing speed.

Like the CPython interpreter, \name overlays Python function calls onto the C
execution stack (each Python call corresponds to a C function call).

Python exceptions are emulated using C++ exceptions.  This allows \name to leverage the builtin
C++ exception handling facility, and greatly simplifies the main interpreter code.  (Proper 
handling of exceptions and return values is a significant source of complexity for the mainline
Python interpreter).

\section{Evaluation}

We evaluated the runtime performance of \name on  a variety of benchmarks.  All
tests were performed on a machine with 8GB of memory and a Xeon W3520
processor.  For most benchmarks, \name provides a small performance benefit
over the CPython interpreter.  For benchmarks that are bound by loop or
interpreter dispatch overhead, \name is over twice as fast as CPython.

\begin{figure}[h!]
\centering
\begin{tabular}{|c|c|} \hline
  \textbf{Benchmark}                   & \textbf{Description}  \\
  \hline Matrix Multiplication       & Multiply square matrices         \\
  Decision Tree               & Recursive control flow           \\
  Wordcount                   & \# of distinct words  \\
  Crypto                      & AES encrypt+decrypt              \\
  Quicksort                   & Classic sorting algorithm        \\
  Fasta                       & Random string generation         \\
  Count threshold             & \# values $>$ threshold \\ 
  Fannkuch\cite{fannkuch}     & Count permutations  \\ 
  \hline
\end{tabular}
\caption{Benchmark Descriptions} \label{tab:benchmarks} 
\end{figure}

\begin{figure}[h] \centering \includegraphics[width=3in]{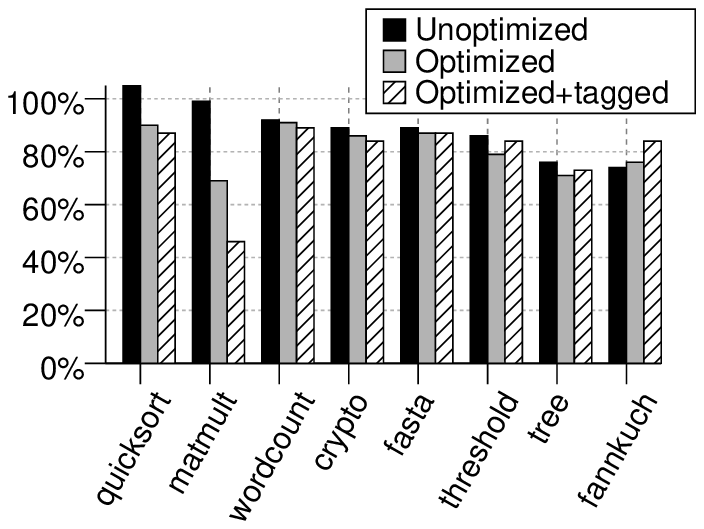}
\caption{\name performance relative to Python\label{fig:runtime-perf}}
\end{figure}

Figure~\ref{fig:runtime-perf} shows the runtime performance of \name relative
to the runtime of CPython interpeter.  The three bars represent the time taken
by (1) unoptimized \name code using untagged (ordinary PyObject) registers, (2)
optimized code with untagged registers and (3) optimized code with tagged.
registers. We were surprised by the inconsistency of benefit from using tagged
registers.  For some benchmarks, such as matrix multiplication, the performance
improvement from switching to tagged registers was quite dramatic.  Most of the
other benchmarks saw either little improvement or even some slowdown from the
switch.

\begin{figure}[h] \centering \includegraphics[width=3in]{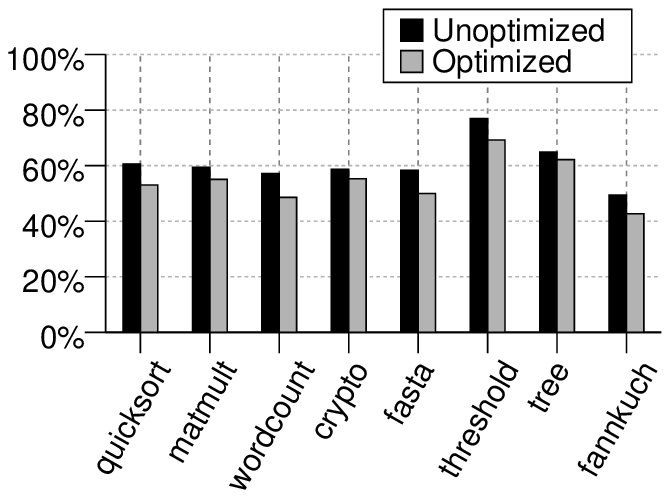}
\caption{Effect of compiler optimizations on number of opcodes, relative to
number of Python stack operations\label{fig:compiler-opcode}}
\end{figure}

We also looked at the change in the number of instructions used after converting
to \rcode, and after optimizations have been run.
(figure~\ref{fig:compiler-opcode})  As expected, the \rcode version of each
benchmark requires significantly fewer instructions to express the same
computation, using on average 45\% fewer instructions.

A few interesting observations can be made:

\begin{itemize}
\item \textbf{Compile times are negligible.} All of our benchmark results include time
taken to compile from stack to \rcode and run the optimization passes.  Despite
our best efforts at making an inefficient compiler (copies used many places
where references might suffice, multiple passes over the code), the time taken
to convert and optimize functions is very small, varying from 0.1ms for simple
functions to 1.1ms for the most complex function in our benchmark set (AES
encryption).  This implies that is profitable to simply convert everything to
\rcode, rather than relying on profile-based techniques (such as
those used by Psyco~\cite{rigo04}) to determine whether it is worthwhile.

\item \textbf{Optimization is important.}  For some benchmarks, the compiler
optimizations result in a 30\% improvement over the unoptimized code; in some
cases changing \name from being slower than CPython to being significantly
faster.  The register code is more amenable to optimization, but the register
machine instructions are slower and more expensive to dispatch than simple stack
machine operations.

\item \textbf{Bit tagging registers yields mixed results.}  Switching to a more
compact and efficient internal representation seemed like it would be a
straightforward win, but this is not always the case.  The potential benefit of
using a tagged inline integer value must be weighed against the potential cost
of converting these integers into Python objects whenever they must be passed
into API functions. In functions that are dominated by arithmetic and logic
operations, tagged registers are a performance win. In other functions,
however, unpacking an integer value to be stored directly in a register is
simply wasted work. 

\end{itemize}

\section{Related Work}
Many different projects have sought to speed up the performance of Python
programs using a variety of techniques.

Nuitka, Cython~\cite{cython}, and ShedSkin reduce the runtime overhead of opcode
dispatch by statically compiling Python programs into C API calls. This
approach, if combined with aggressive optimization, can also remove
some redundant runtime checks. The disadvantage of this approach is that it
requires an explicit (and sometimes lengthy) compilation step, which is at odds
with the usual programming style in a dynamic language like Python.

The currently most popular approach to accelerating dynamic language is tracing
just-in-time (JIT) compilation ~\cite{gal07}, which has proven particularly
effective for JavaScript~\cite{gal09}. One of the primary ways a JIT is able to
achieve good performance is by using unboxed representations for data, which is
incompatible with a native API that exposes the internal representation of data.
Unfortunately, this is the case with Python.  The only currently active JIT
project for Python is PyPy~\cite{bolz09}.  Although PyPy is able to achieve
impressive performance gains, it does so at the expense of breaking C API
compatibility.  This is particularly problematic for scientific Python
libraries, which act largely as wrappers over pre-compiled C or Fortran code and
are often written with particular expectations about the Python object layout.

Psyco~\cite{rigo04} was an older (now abandoned) just-in-time compiler for
Python. By coupling intimately with the Python interpreter and switching between
efficient (unboxed) representations and externally compatible boxed
representations, Psyco was able to avoid breaking C extensions. Unfortunately,
this compatibility required a great deal of conceptual and implementation
complexity, which eventually drove the developers to abandon the project in
favor of PyPy.

\section{Conclusion}

To investigate how fast we could make a binary compatible interpreter for
Python, we built \name, a fast register based compiler and virtual machine for
Python.  \name combines many well-known techniques and few new ones in order to
achieve a significant speedup over regular Python.

What did we learn from the experience?

\textbf{Stack and register bytecodes aren't too different.} Our register based
interpreter proved to be \ntilde25\% faster then the basic Python stack interpreter for
most tasks.  While this is a nice improvement, much larger gains could be made
if we had the ability to change the object format.

\textbf{Tagged object formats are important.} The performance improvement of
using an inline tagged format (integer or NaN tagging) for primitive types is
worth the extra effort; for any sort of performance sensitive code, it easily
means the difference between an interpreter that is 5 times slower then C and
one that is 100 times slower.  If this type of object format could be used
uniformly within the CPython interpreter, it would greatly improve the
performance for almost every task.

\textbf{API design.} The most important lesson we can draw from our experience
is that interpreter APIs should be designed with care. In particular, an API
which exposes how the internals of an interpreter work may be convenient for
gaining a quick performance boost (i.e. use a macro instead of a function), but
in the long-term, exposing these internal surfaces makes it nearly impossible to
change and improve performance in the future.  For Python, it is not the size of
the C API that is the problem, but rather its insistence on a particular object
format.

The assumption made by an API are not always obvious.  For instance, when
writing an API for an interpreter, it may be tempting to have functions which
directly take and return object pointers.  This simple decision has unexpected
consequences; it prevents the use of a copying garbage collector.

\section{Future Work}

One of our goals with \name was to build a platform that would simplify writing
new experiments.  The use of \rcode and a pass based compiler format makes
trying out new optimization techniques on Python bytecode easy.  Particular
ideas we would like to explore in the future include:

\begin{itemize}
\item \textbf{Type specialization}. At compile time, type propagation
can be performed to determine the types of registers.  Unboxed, type specific
bytecode can then be generated to leverage this information.

\item \textbf{Container specialization}. The performance benefit of tagged registers is
primarily limited by the need to convert to and from the CPython object format
whenever an API call is made.  This is almost always due to a register being
stored into a Python list or dictionary object.  We can improve on this by 
creating specialized versions of lists and dictionaries for each primitive type.
These specialized objects would support the standard list/dictionary interface
and convert to and from the Python object format on demand (thus allowing them 
to be used in external code); internally they would store objects in an
efficient tagged format.

\item \textbf{Improving attribute hints}. The current \name hinting mechanism improves
performance slightly, but is very limited in its application.  Better results
could be obtained by making lookup more explicit in the bytecode (first check
for accessor functions, then look up the actual name).
\end{itemize}

The source code for \name is available online at:\\
\url{http://github.com/rjpower/falcon/}; we encourage anyone who is interested to try it
out and provide feedback.

\bibliographystyle{acm}
\bibliography{ref}

\begin{thebibliography}{10}

\bibitem{cython}
{C}ython: {C}-{E}xtensions for {P}ython.
\newblock \url{http://cython.org/}.

\bibitem{gcc-labels-as-values}
{GCC} documentation: Labels as {V}alues.
\newblock \url{gcc.gnu.org/onlinedocs/gcc/Labels-as-Values.html}.
\newblock Accessed: August 13th 2013.

\bibitem{python-token-threading}
Python 3 token threading.
\newblock \url{bugs.python.org/issue4753}.
\newblock Accessed: August 13th 2013.

\bibitem{python-api}
{P}ython/{C} {API} {R}eference {M}anual.
\newblock \url{docs.python.org/2/c-api/}.
\newblock Accessed: August 13th 2013.

\bibitem{fannkuch}
{\sc Anderson, K.~R., and Rettig, D.}
\newblock Performing {Lisp} analysis of the {FANNKUCH} benchmark.

\bibitem{dynamo}
{\sc Bala, V., Duesterwald, E., and Banerjia, S.}
\newblock Dynamo: a transparent dynamic optimization system.
\newblock In {\em {PLDI} 2000\/} (2000), vol.~35, ACM, pp.~1--12.

\bibitem{direct-threading}
{\sc Bell, J.~R.}
\newblock Threaded code.
\newblock {\em Commun. ACM 16}, 6 (June 1973), 370--372.

\bibitem{bolandmemory}
{\sc Boland, C.}
\newblock {\em Memory Allocation and Access Patterns in Dynamic Languages}.
\newblock PhD thesis, Heinrich Heine University Düsseldorf, 2012.

\bibitem{bolz09}
{\sc Bolz, C.~F., Cuni, A., Fijalkowski, M., and Rigo, A.}
\newblock Tracing the meta-level: {PyPy}'s tracing {JIT} compiler.
\newblock In {\em Proceedings of the 4th workshop on the Implementation,
  Compilation, Optimization of Object-Oriented Languages and Programming
  Systems\/} (2009), ACM, pp.~18--25.

\bibitem{abstract-interpretation}
{\sc Cousot, P., and Cousot, R.}
\newblock Abstract interpretation: a unified lattice model for static analysis
  of programs by construction or approximation of fixpoints.
\newblock In {\em Proceedings of the 4th ACM SIGACT-SIGPLAN symposium on
  Principles of programming languages\/} (1977), ACM, pp.~238--252.

\bibitem{ssa}
{\sc Cytron, R., Ferrante, J., Rosen, B.~K., Wegman, M.~N., and Zadeck, F.~K.}
\newblock Efficiently computing static single assignment form and the control
  dependence graph.
\newblock {\em ACM Transactions on Programming Languages and Systems 13}, 4
  (Oct 1991), 451--490.

\bibitem{davis03}
{\sc Davis, B., Beatty, A., Casey, K., Gregg, D., and Waldron, J.}
\newblock The case for virtual register machines.
\newblock In {\em In Interpreters, Virtual Machines and Emulators (IVME '03)\/}
  (2003), ACM Press, pp.~41--49.

\bibitem{gal07}
{\sc Gal, A., Bebenita, M., and Franz, M.}
\newblock One method at a time is quite a waste of time.
\newblock In {\em Proceedings of the Second ECOOP Workshop on Implementation,
  Compilation, Optimization of Object-Oriented Languages, Programs and
  Systems\/} (2007).

\bibitem{gal09}
{\sc Gal, A., Eich, B., Shaver, M., Anderson, D., Mandelin, D., Haghighat,
  M.~R., Kaplan, B., Hoare, G., Zbarsky, B., Orendorff, J., Ruderman, J.,
  Smith, E.~W., Reitmaier, R., Bebenita, M., Chang, M., and Franz, M.}
\newblock Trace-based just-in-time type specialization for dynamic languages.
\newblock In {\em PLDI\/} (2009), pp.~465--478.

\bibitem{polymorphic-inline-caching}
{\sc H{\"o}lzle, U., Chambers, C., and Ungar, D.}
\newblock Optimizing dynamically-typed object-oriented languages with
  polymorphic inline caches.
\newblock In {\em ECOOP'91 European Conference on Object-Oriented
  Programming\/} (1991), Springer, pp.~21--38.

\bibitem{numpy}
{\sc Oliphant, T.~E.}
\newblock Python for scientific computing.
\newblock {\em Computing in Science \& Engineering 9}, 3 (2007), 10--20.

\bibitem{rigo04}
{\sc Rigo, A.}
\newblock Representation-based just-in-time specialization and the {Psyco}
  prototype for {Python}.
\newblock In {\em Proceedings of the 2004 ACM SIGPLAN symposium on Partial
  evaluation and semantics-based program manipulation\/} (2004), ACM,
  pp.~15--26.

\bibitem{showdown08}
{\sc Shi, Y., Casey, K., Ertl, M.~A., and Gregg, D.}
\newblock Virtual machine showdown: Stack versus registers.
\newblock {\em ACM Trans. Archit. Code Optim. 4}, 4 (Jan. 2008), 2:1--2:36.

\bibitem{python}
{\sc Van~Rossum, G., et~al.}
\newblock Python programming language, 1994.

\end{thebibliography}

\end{document}